RESEARCH ARTICLE

# Deep fusion of gray level co-occurrence matrices for lung nodule classification

Ahmed Saihood[1,2], Hossein Karshenas[1]*, Ahmad Reza Naghsh Nilchi[1]

1 Artificial Intelligence Department, Faculty of Computer Engineering, University of Isfahan, Isfahan, Iran,
2 Faculty of Computer Science and Mathematics, University of Thi-Qar, Nasiriyah, Thi-Qar, Iraq

* h.karshenas@eng.ui.ac.ir

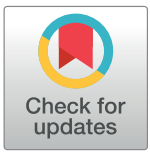

## Abstract

Lung cancer is a serious threat to human health, with millions dying because of its late diagnosis. The computerized tomography (CT) scan of the chest is an efficient method for early detection and classification of lung nodules. The requirement for high accuracy in analyzing CT scan images is a significant challenge in detecting and classifying lung cancer. In this paper, a new deep fusion structure based on the long short-term memory (LSTM) has been introduced, which is applied to the texture features computed from lung nodules through new volumetric grey-level-co-occurrence-matrices (GLCMs), classifying the nodules into benign, malignant, and ambiguous. Also, an improved Otsu segmentation method combined with the water strider optimization algorithm (WSA) is proposed to detect the lung nodules. WSA-Otsu thresholding can overcome the fixed thresholds and time requirement restrictions in previous thresholding methods. Extended experiments are used to assess this fusion structure by considering 2D-GLCM based on 2D-slices and approximating the proposed 3D-GLCM computations based on volumetric 2.5D-GLCMs. The proposed methods are trained and assessed through the LIDC-IDRI dataset. The accuracy, sensitivity, and specificity obtained for 2D-GLCM fusion are 94.4%, 91.6%, and 95.8%, respectively. For 2.5D-GLCM fusion, the accuracy, sensitivity, and specificity are 97.33%, 96%, and 98%, respectively. For 3D-GLCM, the accuracy, sensitivity, and specificity of the proposed fusion structure reached 98.7%, 98%, and 99%, respectively, outperforming most state-of-the-art counterparts. The results and analysis also indicate that the WSA-Otsu method requires a shorter execution time and yields a more accurate thresholding process.

## Introduction

One of the primary challenges in human life is the prevalence of fatal diseases. Today, lung cancer has the highest mortality rate [1]. Studies reveal that the number of patients diagnosed with lung cancer has increased by approximately 25% from 2005 to 2018 [2]. Researchers worldwide have noticed that lung cancer rates are gradually rising, potentially threatening human health. The early discovery of lung cancer is vital as it increases the probability of cure and survival [3].

Cancer aggressively divides lung cells, and cancerous cells continuously increase inside the lung. As the disease progresses, the damaged lung textures become more extensive. Cancer cells can propagate to other body parts and intensify the severity of the disease. Lung cancer development can be divided into four stages based on the nodule size and the growth of the damaged area in the lung (Fig 1). In the first stage, the cancerous mass constitutes less than 3 cm of the lung texture. The bronchi are engaged in the second stage. In the third stage, the trachea and the upper lope sections of the lung are affected by cancer cells. In the last stage, both lung lobes become involved, and the probability of death is high. In this stage, cancer can extend within the lungs or to an area outside the lungs [4].

Image-based computer-aided diagnosis (CAD) systems are applied to detect and diagnose lung cancer by analyzing CT scan images automatically, specifically their textures. In CAD systems, image processing has the following stages: 1) noise removal from the image; 2) feature extraction; 3) feature selection; 4) cancer segmentation, and 5) cancer classification [5, 6].

Most methods for automatic lung cancer detection and classification combine image processing techniques with deep learning and machine learning methods. Many methods have been proposed, including support vector machines (SVM) [7], artificial neural networks (ANN) [8], random forests [9], genetic algorithms (GA) [10], fuzzy C-means clustering [11], and deep learning-based features fusion [12].

Simple traditional segmentation methods such as the similarity-based segmentation techniques [13], multi-shape graph-cut approach [14], region aided geometric snakes approach [15], and feature-based atlas approach [16] are fast but do not meet the segmentation accuracy required for lung CT images with complicated grey level patterns. The loss of quality in CT imaging due to the risk of ionizing rays [17] is one of the major challenges in this regard. Regional deconvolutional neural networks for image segmentation [18] seek to overcome this drawback at the expense of extended run-time. The continuous expansion of CT scanning techniques and their application in automatic CAD systems have made segmentation methods

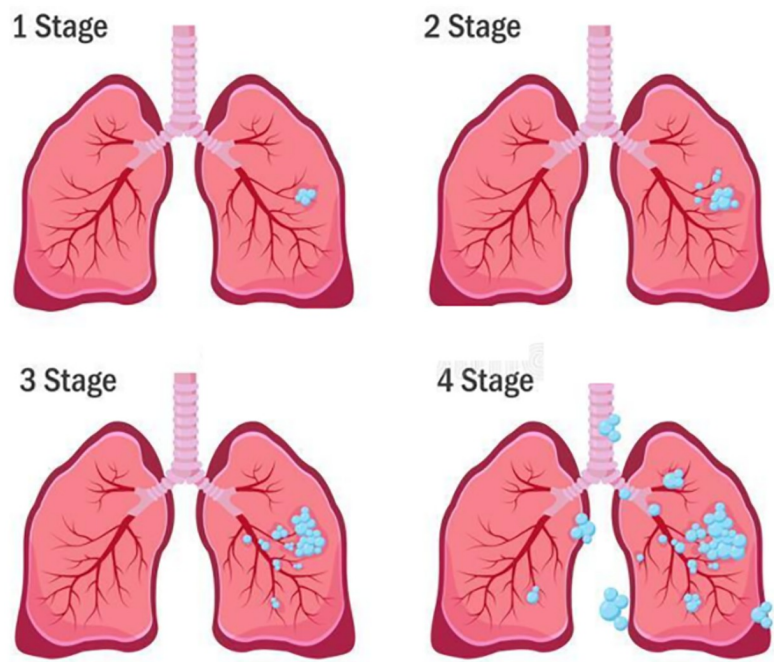

**Fig 1. The four stages of lung cancer development.**

https://doi.org/10.1371/journal.pone.0274516.g001

with high accuracy and low computational complexity highly desirable. A well-established and efficient image segmentation approach for detecting cancerous tissues is Otsu thresholding [19–21]. This method increases the between-class variance through an exhaustive search. However, this process can be time-consuming, depending on the number of thresholds (i.e., the segments) sought. Many methods are proposed to enhance the original Otsu method by preprocessing the input image before segmentation [19] or applying the Otsu algorithm iteratively to the regions of interest in the image [22]. Intelligent meta-heuristic search, such as particle swarm optimization (PSO) algorithm, can be used to find the proper thresholding values among multi-thresholds of Otsu segmentation [23]. The problem with PSO is its low convergence rate; it cannot find proper solutions in a limited time. In this study, the water strider optimization algorithm (WSA) is applied to search for thresholding values during the non-small cell lung cancer (NSCLC) detection phase for a faster and more accurate segmentation of lung CT scans by avoiding outliers. WSA has a good convergence rate and obtains reasonable solutions in a limited time when combined with the Otsu segmentation method, resulting in a more accurate detection of lung nodules. After detection, features are extracted with a far less computational effort from the regions of interest in lung CT images.

Convolutional neural networks (CNNs) have been broadly applied in lung image processing and classification [24] due to their well-known representation learning capabilities. However, the volumetric shape of lung nodules is captured in a sequence of CT scan slices. The arrangement of adjacent slices in the CT scan allows the information in a slice to be related to the next in a spatial sense. Feed-forward CNN architectures can face problems extracting discriminative features from this type of data, especially with many slices in a CT scan. Moreover, contrast loss and image blurring make the medical images unclear and their adjustment in 3D feature extraction difficult. On the other hand, CNN models cannot encode the location and orientation of an object [25] and require fixed-size input images. Therefore, raw images must be resized to the smallest image when the problem has inputs of different sizes. Because of this, usually, 3D-CNNs are applied by dividing the input image into a stack of patches to solve the fixed-size problem [26, 27]. Due to these inherent complexities, lung cancer classification has not yet achieved the desired accuracy.

Gray level co-occurrence matrices (GLCMs), first proposed by Haralick et al. [28], provide a good description of image texture features. These co-occurrences are computed for a grey-level image by enumerating the frequency of pixel pair values. Such spatial information, extracted in different directions for each pixel, can be used to improve representation learning in CNN-based architectures [29]. Also, the fusion of specific Haralick features with other handcrafted and in-depth features extracted through deep learning models has been used, revealing a high impact on the lung cancer classification performance [30, 31]. Although this joint consideration of features provides better representation performance, its impact on learning high-level features is low [32] because of existing unrelated or redundant features, ignoring the spatial distribution of texture within the volumetric shape, and the relationships between nodule's features (i.e., long-range dependencies in the texture information extracted from the slices).

This article proposes a new 3D-GLCM computation approach to overcome these drawbacks when considering the volumetric shape of lung nodules. In this method, the CT scan is divided into several volume spaces (VSs), and a 3D-GLCM is computed for each in 13 directions instead of performing the 3D-GLCM computations for the whole CT scan at once, which is computationally very expensive. The local 3D-GLCMs computed from different VSs would allow approximating the distribution of textural information in volumes, thus providing an important source of information for classification. Moreover, by reducing the computational complexity of obtaining local GLCMs, we can increase the resolution of quantized values considered in the computations, resulting in more discriminative texture features. The fusion of these local VS-based 3D-GLCMs allows the classification model to utilize the correlation between sequenced slices in the whole nodule volume.

The traditional fusion algorithms based on CNNs focus on a one-batch training mode to estimate the model's parameters instantly from the whole data. However, when applied to sequences of data (e.g., slices in a CT scan), this training strategy is greatly affected by the limitation in data, specifically with the existence of redundant or irrelevant information. Recurrent neural networks (RNNs) [33] can effectively process long sequential information using feedback connections between several units. Long short-term memory (LSTM), as one of the important extensions of RNN, has been used in this paper to design a deep neural architecture for the fusion of local GLCMs and to classify lung nodules into three classes, benign, malignant, and ambiguous. Furthermore, an approximation of the 3D-GLCM computation with 2.5-D-GLCM is performed, where the co-occurrences are extracted from every slice in eight directions throughout the whole CT scan and then fed to the LSTM-based fusion structure. The proposed fusion and classification architecture has also been compared with other CNN-based architectures using 2D-GLCMs and raw CT scan slices.

The main challenges addressed in the proposed method are: 1) accurate and fast detection of nodule regions within slices; 2) extraction of high-level GLCM-based features for nodule representation; and 3) effective fusion of the extracted sequential features for nodule classification.

This study aims to detect and classify lung nodules accurately and determine their cancerous or benign nature. The sequential structure of CT scans is promising for the volumetric processing of nodules in lung images. The proposed method consists of 1) CT scan preprocessing, 2) nodules detection, 3) GLCM computations, 4) LSTM-based deep fusion structure, and 5) nodule classification.

The main contributions of this article are:

1. A new LSTM-based deep neural fusion structure, where long-range dependencies are handled through texture fusion for lung nodule classification;

2. A new method of volumetric GLCM computation, where more spatial relationships among nodule voxels can be captured through specific volume spaces (VSs), reducing the unnecessary complexity and improving the fusion model performance;

3. An extended Otsu-based segmentation method that, in combination with WSA, overcomes the existing drawbacks regarding time consumption and thresholding enhancement.

The rest of article is organized as follows. The literature is reviewed in the Related studies section, the proposed method is introduced in the Materials and methods section, the design is elaborated in the Experimental design section, results and discussion are presented in the Results and discussion section, and the article is concluded in Conclusions and future work.

## Related studies

Deep learning and machine learning algorithms have been applied in various studies to detect and classify lung nodules. The noise in the images and the orientation of nodules make the detection process complex and exhaustive. Pan et al. [34] introduced a double-path convolutional neural network (DPCNN) to reduce such complexities. It consists of a depletion and detection method in an end-to-end approach. First, a learning model is adopted for noise reduction in the preprocessing step. Then, the DPCNN is applied as a feature extractor into double paths as the input for lung cancer detection. One path applies MobileNetV2 [35], and the other uses a more massive input image but with short well-proportioned layers. This method needs a long learning time and does not provide a mechanism to improve segmentation. Nurtiyasari et al. [36] applied CT scan images to extract and preprocess histogram features based on local energy to detect lung cancer. Their proposed method was to improve the detection by applying preprocessing and local energy histograms to enhance the resolution and quality of the obtained images.

Deep learning techniques are used to find fused outputs from the training process and incorporate them into the analysis. Hemant et al. [37] applied the primary preprocessing method, where the Local Energy-Based Shape Histogram technique was used for better results. In the proposed method, the medium filters are applied to smoothen the images, improve their quality, and segment them through segmentation techniques. Priyadharshini et al. [38] developed a lung cancer detection method by a CNN combined with the Bat Algorithm. They sought to detect cancer nodules in the lungs by the input image of the lungs and the structure of lung cancer formation. Features are extracted by discrete wavelet transform. A fuzzy C-means clustering method is used to determine lung cancer nodules, and a modified CNN algorithm based on the bat algorithm was introduced to classify lung cancer effectively. The analysis by Deepti [39] revealed the essence of machine learning algorithms in predicting and detecting different cancer types. Riquelme et al. [40] assessed the advanced learning algorithms and architectures for automated detection systems applied in lung cancer detection. The existing lung CT scan datasets are compared with different techniques in detail. Kriegsmann et al. [41] used deep learning to classify small and non-small lung cancer cells. A group of lung cancer images is collected for processing and assessment, and using an optimized architecture of InceptionV3 CNN with the highest classification accuracy the test set is classified. A 95% accuracy for detecting patients with cancer is reported in the test set. This method highlights the potentials and limitations of CNN-based image classification models in discriminating lung tumors.

Bhandary et al. [42] reviewed a deep learning framework for detecting lung abnormalities and found that their early detection reduces the risk through fast and efficient treatment. A deep learning framework is used to analyze pneumonia and lung cancer. Two deep learning solutions are considered in assessing the problem: 1) the modified AlexNet for classifying CT scan images of the chest, where classification is made through an SVM, and 2) using a mixture of handcrafted and learned features from deep learning to improve classification accuracy during lung cancer assessment. The feature extraction based on principal component analysis can enhance the feature vector.

Many studies have explored methods of classifying lung cancer by 3D-CNN based on CT scan images. Pradhan et al. [43] applied a series of morphological processes to remove the lung nodules mask and introduce the detected objects to the 3D-CNN model. Khumancha et al. [44] applied cubic masks to specify the region of interest (ROI) in CT scans and presented the predicted regions to 3D-CNN for lung cancer detection. 3D-CNN has a prolonged training time, and most CNNs models fail to recognize noisy images, leading to irrelevant feature extraction. Wang and Chakraborty [45] enhanced the efficiency of lung cancer classification training by applying an average pooling to input nodule volumes fed to RNN. The spatial correlation in multiple slices representing nodules is ignored in their work. Nurtiyasari et al. [36] applied a wavelet RNN model to remove the noise in the input raw images. RNN classifies the images into cancerous and non-cancerous.

Several researchers have classified lung nodules based on GLCM texture features. Alotaibi [46] proposed a GLCM-based neural network model that involved a preprocessing phase of extracting the regions of interest from slices through GLCM computations. Then, the extracted features are provided for the radial bias function neural networks for classification. Thohir et al. [47] provided texture features extracted from medical images by applying GLCMs to an SVM classifier. They concluded that a polynomial kernel achieves better results. Tan et al. [48] calculated the 3D-GLCM in 13 directions from volumetric CT scans and presented it to CNN. The slices were first scaled from the Hounsfield units to 32 grey levels, and then the GLCM was computed for every CT scan and fed to a CNN in a different channel. The CNN might miss some correlated information from adjacent slices in the CT scan.

Feature representation can be biased for a particular learning model. De la Torre [49] applied multi-modal feature learning frameworks to overcome this drawback by combining features from multi-modal, leading to high-dimensional data and posturing a problem in the final pattern recognition step. The combined feature vectors can contain redundant ingredients and noise. Zhang et al. [50] proposed a hierarchical subnetwork-based neural network to overcome this issue and degrade the dimension of the feature vector. They introduced an iterative learning process rather than stacking separate blocks.

The hybridization of deep neural networks is another popular research subject. Muthusamy et al. [51] proposed a CNN-RNN model, where the CNN extracted patterns from the images, and RNN is applied to classify the images. They applied a four-level discrete wavelet transform to disband the image into different sub-bands. The texture features are extracted through GLCM. Jiang et al. [52] used a 3D-convolution-LSTM architecture to classify lung nodules where the input CT scans were first fed into a convolution layer, and LSTMs are then deployed to detect the nodule mask. Mhaske et al. [53] proposed a CAD system containing CT scan segmentation to detect lung nodules through Otsu thresholding, features extraction by CNN, and classification by RNN-LSTM.

Recently, researchers have considered the lung nodule variation in size and location through their slices. Peng et al. [54] proposed 3D multi-scale deep convolution neural networks. The networks used the Res2Net module to capture multi-scale feature information from the nodule and a region proposal network to improve the contextual and spatial attention maps. Al-Shabi et al. [55] used a non-local attention mechanism to resolve the long-range dependency for the volumetric lung nodules. They proposed a progressive channel attentive non-local network for lung nodule classification. In addition to 3D axial attention map extraction, a channel attention mechanism captures the channel's attention.

Studies on improving image segmentation for lung cancer detection are growing. Yu et al. [56] applied the K-mean method to inspect the similarity of pixels by dividing them into subgroups. These data are classified through deep learning with instantaneously trained neural networks. Chithra and Roy [20] adopted the Otsu thresholding method for detecting lung nodules. Vas and Dessai [57] performed morphological operations on CT scans to remove the foreground edges. The closing procedure incorporated a disk structural element to reveal the lung mask, and a median filter is then applied to process the internal features of the lung. Nizami et al. [58] applied the K-mean clustering to isolate the lung tissue. Wavelet packets containing five sub-bands with the most considerable energy and entropy are used to cluster coefficients. Park and Guldmann [6] applied median and average filters to enhance the input images for thresholding through Otsu. Helen et al. [19] sought optimal thresholds by combining PSO with Otsu thresholding for CT scan segmentation to avoid the exhaustive search.

## Materials and methods

The method proposed in this study consists of five steps: CT scans preprocessing, nodule detection, GLCM computations, LSTM-based deep fusion structure, and nodule classification (Fig 2).

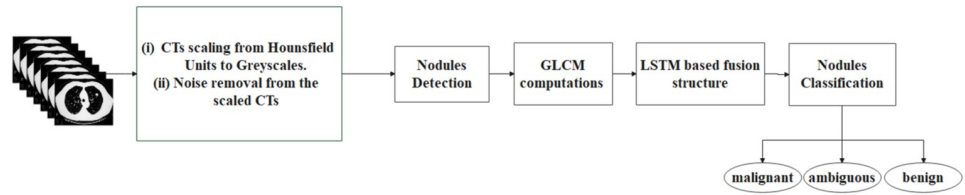

**Fig 2. The constituent steps of the proposed method.**

https://doi.org/10.1371/journal.pone.0274516.g002

In the preprocessing step, the CT scans are rescaled from Hounsfield units to greyscale, and the noise is removed from the images. The WSA-Otsu segmentation method is applied in lung nodule detection. The volumetric GLCM computations are derived from the CT images divided into several VSs. The LSTM-based deep fusion structure combines the computed volumetric GLCMs for the adjacent VSs, and the lung nodules are classified into benign, malignant, and ambiguous.

### 3D-GLCM computations

The nodule volume is divided into many sub-volumes (VSs) for the proposed 3D-GLCM computations. To consider the volumetric shape of nodules, the sequence of 3D-GLCMs computed from these sub-volumes are fed to an especially adjusted LSTM-based deep fusion structure for fusion. The 3D voxels co-occurrences in 13 directions are considered for computing 3D-GLCMs, instead of 2D pixels in the slices. More directions may add more information, leading to unnecessary complexity and possible redundancy in the texture data. The fusion of adjacent VSs allows the model to incorporate more spatial information between slices in CT scans. The GLCMs are computed for each VS by combining the adjacent 3D-GLCMs through the proposed LSTM-based deep fusion structure.

### Deep recurrent structure for fusion

The structure of the proposed 3D-GLCMs fusion model is shown in Fig 3. The proposed LSTM layer fuses the 3D co-occurrence features extracted from each VS with those of the adjacent VSs. The slices of CT scan X are divided into $p$ VSs, and computations are performed for each VS to obtain the corresponding 3D-GLCM. The obtained 3D-GLCMs are presented to $p$ LSTM units, and the representation of the spatial information from the features extracted from the adjacent VSs is adjusted internally. In this way, the contextual information processed in one LSTM unit (that is, the features extracted from one VS), is transferred to the following LSTM unit.

### Hybridization of Otsu and WSA

The WSA is combined with the Otsu thresholding technique to enhance image segmentation. In the first phase of the proposed method, several random threshold vectors as female and male insects are established as members of the water strider (WS) population. In phase two, a male solution mate with a probability of $P$ (Eq (1)) for updating the thresholds using the mating process.

$$X_i^{t+1} = \begin{cases} X_i^t + R.rand & mating \\ X_i^t + R.(1 + rand) & nomating \end{cases} \quad (1)$$

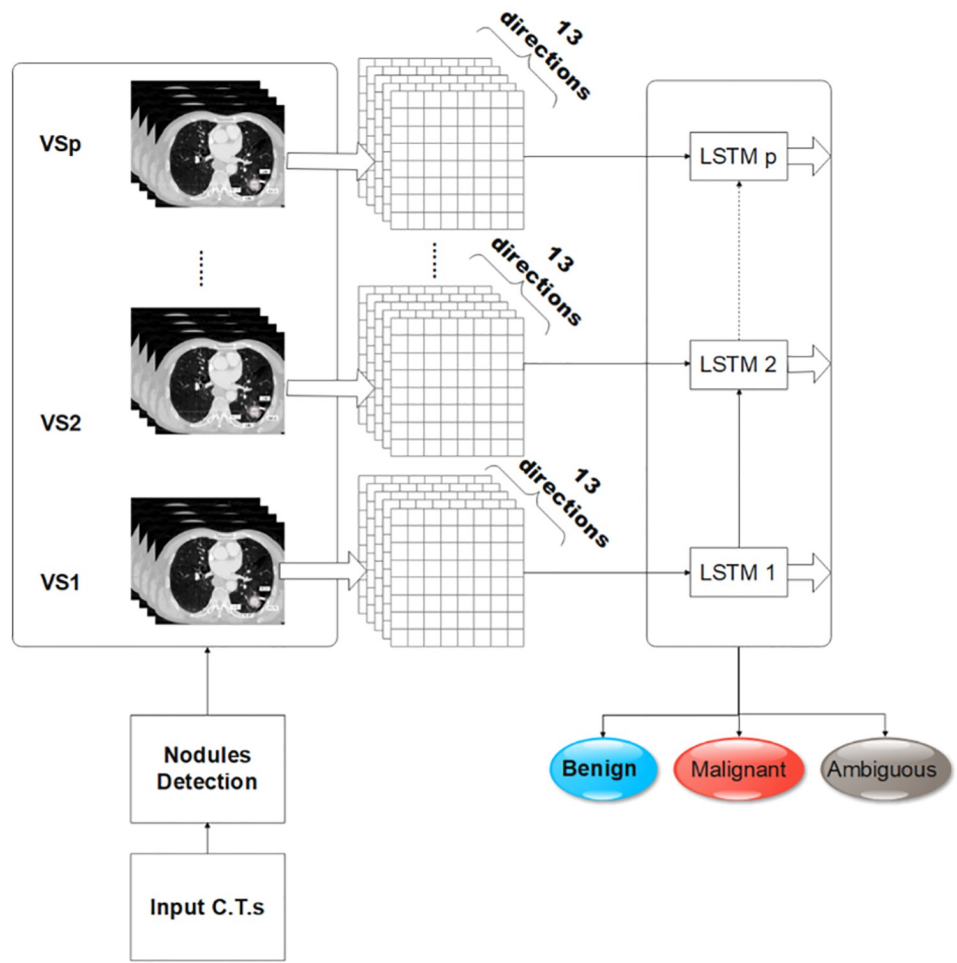

**Fig 3. The 3D-GLCMs are computed from VSs of nodule volume and fed to a recurrent neural network.**

https://doi.org/10.1371/journal.pone.0274516.g003

where $X_i^{t+1}$ is the position of WS or an Otsu thresholding vector at iteration $t + 1$, $X_i^t$ is the location of the $i^{th}$ WS at step $t$, $r$ is a random number in the [0, 1] range, and $R$ is the distance between a male WS $X_i^t$ and a female $X_F^t$, calculated by:

$$R = ||X_F^t - X_i^t|| \quad (2)$$

When the foraging behavior of water striders is considered, the Otsu thresholds can be updated by Eq (3).

$$X_i^{t+1} = X_i^t + 2rand.(X_{BL}^t - X_i^t) \quad (3)$$

A WS searches the space between itself and the best solution to find a better solution. If a solution is not improved after foraging, it is moved to another territory for more foraging (getting closer to the best solution). If the quality of the solution is still low, then it is replaced by a new random WS larva inside the source territory, according to Eq (3). The best thresholds at each iteration of the WSA are extracted by applying this optimization approach to the Otsu thresholding algorithm. The last iteration contains the best thresholds applicable to image segmentation. The maximized objective function is defined when $p_i$ is the proportion of pixels with an *i* intensity, obtained through Eq (4).

$$p_i = \frac{n_i}{N}, p_i \geq 0, \sum_{i=0}^{L} p_i = 1 \qquad (4)$$

In this equation $n_i$ is the frequency of pixels in the image with an intensity of *i*, and *N* and *L* are the total pixel count and the grey level count in the image, respectively. For thresholding images into multiple classes ($c_1, c_2, \ldots c_m$), *m* is the threshold count, and WSA is the population of threshold vectors. In Otsu thresholding, the sum of intensity frequencies in each class (the mean proportion of grey levels in a class) is obtained through Eq (5).

$$w_k = \sum_{i \in C_k} p_i \qquad (5)$$

The weighted average of intensities based on the considered thresholds (cumulative probability) is obtained through Eq (6). In the next step, the average intensity in all classes for the whole image is obtained through Eq (7).

$$\mu_k = \sum_{i \in C_k} \frac{i \cdot p_i}{\omega_k} \qquad (6)$$

$$\mu_T = \sum_{k=1}^{m} \omega_k \mu_k \qquad (7)$$

In this equation $\mu_k$ is the average intensity weighted in the $k^{th}$ class, and $\mu_T$ is the average image intensity. The objective function is defined through Eq (8) to obtain the maximum value [59].

$$fitness = \sum_{k=1}^{m} \omega_k (\mu_k - \mu_T)^2 \qquad (8)$$

The Pseudo-code of the hybridization process is shown in Algorithm 1.

## Experimental design

Experiments are performed to assess the proposed methods for lung cancer detection and classification. The details of the LSTM-based deep fusion structure with different GLCM computation modes and are presented here, and their results are reported in the next section. TensorFlow software version 2.4.0 is used for all implementations. The experiments are performed in a system with a Windows operating system and a GEFORCE RTX 2080 GPU.

## Preprocessing

Noise removal is an important operation for segmentation and features extraction preprocessing. To this end, the average or median filter methods are usually adopted. This study adopts the median filter as it introduces no new noise intensity. The output of the median noise removal method is expressed by the following equation:

$$\hat{I} = \underset{((i) \in K)}{\text{median}} \{I_i\} \qquad (9)$$

where *K* is the pixel count adjacent to the central pixel, and $I_i$ is the input image with an $\hat{I}$ value after noise removal.

**Algorithm 1** Pseudocode for WSA-Otsu Segmentation

```
Input Number of Maximum Iterations, 2D Image, Number of territories
      NT, and Number of classes T
Output The segmented 2D image.
1: Initialize the population of WSs with randomly generated vectors of
   size T – 1 threshold as the WS population members.
2: Calculate the objective value for each WS through Eq (8).
3: repeat
4:   Allocate WSs to NT regions (territories) based on their finesses.
5:   for NT do
6:     Members are considered female and primarily male. The position
       of every insect is updated through Eq (1)
7:     Evaluate the new position of the WS by applying the Otsu object
       function (Eq (8))
8:     if the objective function is more valuable than the earlier case
       then
9:       recover the new position
10:    else
11:      The new position of the WS is updated to move to a new terri-
         tory containing more food using Eq (3)
12:      The WS is evaluated in the new territory by the Otsu objec-
         tive function.
13:    end if
14:    if The objective function value is still low then
15:      The WS will be replaced by a ripened larva through Eq (2)
16:      The Otsu objective function evaluates the WS larva.
17:    end if
18:  end for
19: until Iterations reach their maximum
20: Segment the image based on the optimal thresholds counted in the
    last iteration.
21: return segmented image
```

## Experimental setup

In the following paragraphs the detail design of the proposed LSTM-based deep fusion structure used for evaluation is provided and discussed.

First, the fusion structure involving GLCMs computations for 2D-slices is established. Each slice in a CT scan is scaled into (0-255) greyscale levels. After nodule detection, the 2D-GLCMs are calculated for each of the eight directions. The interaction between spatial information calculated from each direction is learned by fusing these GLCMs in an LSTM-based deep fusion structure. In this approach, each slice is labeled as benign, malignant, and ambiguous. The features extracted from ordered directions from 0˚ to 315˚ represent the spatial context information fused internally in the network LSTM layer. The information is obtained from the first LSTM unit, where the features from the direction 0˚ are processed and passed to the next LSTM unit, where the features from the direction 45˚ are processed. This trend continues, and the information passing from every unit is considered in the subsequent units. Such a mechanism captures the dynamic variations in a contextual state over input features history while fusing these features to determine the correlation between each reference pixel and its neighbors in all directions, which ignores the reality of volumetric nodule shapes.

The second experiment involves the 2.5D-GLCM computations, where the features are extracted from multiple slices. The co-occurrence matrix (CM) coordination is ($x$, $y$, $z$) for each slice, where $z$ is the stacked feature in every direction. Fig 4 shows the fusion structure proposed for 2.5D-GLCMs. The volumetric shape of nodules in CTs inspired researchers to

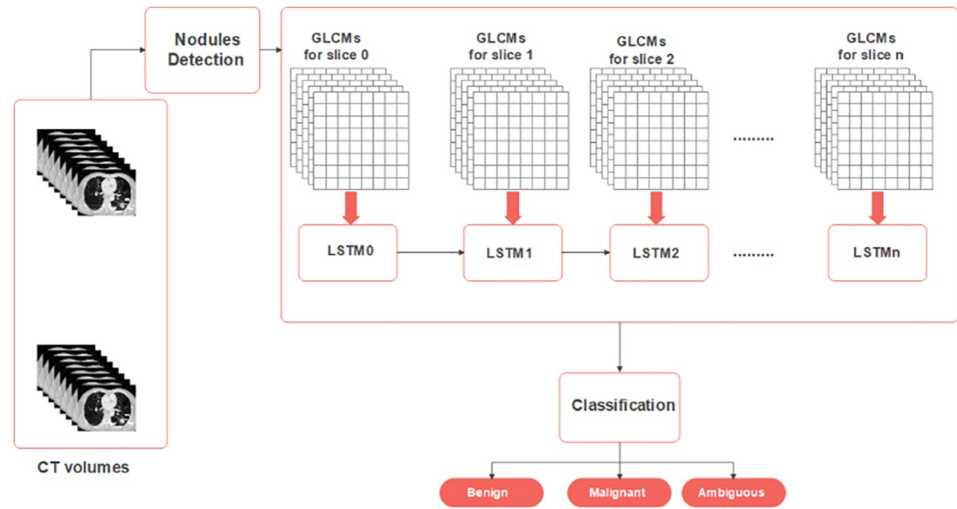

**Fig 4. In 2.5D-GLCM mode the GLCMs computed for each slice are fed to the recurrent neural network.**

https://doi.org/10.1371/journal.pone.0274516.g004

construct this model. The LSTM network fuses the features extracted from different slices by an internal representation of the contextual spatial information among adjacent slices. In this experiment, a richer context is obtained based on the information of not solely each slice but all in a CT scan. The processed features at an LSTM unit in the network pass the contextual state information to the next unit. When fusing these GLCMs through this LSTM-based deep neural architecture, the correlation of adjacent slices gives an advantage over other fusion methods that use CNNs [39], which ignore the spatial information among the slices. Each LSTM unit needs to wait for the previous LSTM unit output and this can cause a long processing time because CT scans consists of long series of slices.

In the third experiment, the proposed LSTM-based deep fusion structure is used to fuse the 3D-GLCM computations, where the nodule volumes are divided into many VSs and then the 3D-GLCMs are computed for the adjacent VSs. Every three sequenced slices are selected as a VS, so that every voxel in the 3D volume VS has a resolution of 1 mm$^3$. In the method using 2D-GLCMs, the CM is calculated for each slice in eight directions, resulting in a $8 \times h \times w$ feature-map, where $h$ and $w$ respectively show the slice's height and width, and fed to the deep LSTM-based neural architecture after vectorization. Thus, eight LSTMs are considered in the input layer of the model. This model contains three LSTMs layers and one dense layer. The vectorized features are fed into eight parallel sequence-to-sequence LSTMs with 128 units. The third LSTM layer is a sequence-to-one that outputs a vector of 128 tensors. Finally, this vector is provided to a dense layer with three units for classification. In this model, the Relu activation function is involved in the LSTM layers and the softmax in the last layer. In the 2.5D-GLCM mode, CMs are calculated for the *n* slices in a CT scan. The features are padded because lung nodules can extend to several slices within the CT scan. Similar to the previous mode, the dimension of the extracted feature map is $8 \times h \times w$ for each slice, and stacking the features in *n* adjusted slices results in a dimensionality of $n \times 8 \times h \times w$, which is fed to the deep network. This model consists of three LSTM and two dense layers. Parallel sequence-to-sequence LSTM units fuse the vectorized elements with 128 units. The last LSTM layer is a sequence-to-one layer with a vector output, fed to a fully connected layer with 32 units. The last dense layer has with three units corresponding to the classification outputs. The Relu activation function is applied in all layers and softmax in the last layer for classification. For 3D-GLCMs, the co-

Table 1. The structure of the proposed LSTM-based fusion method. *f* refers to the number of elements in the extracted features.

| Parameters | 2D-GLCM-LSTM | 2.5D-GLCM-LSTM | 3D-GLCM-LSTM |
|---|---|---|---|
| Input shape | $8 \times f$<br>#*LSTMs*: 8<br>#*Features*: *f* | $n \times f$<br>#*LSTMs*: *n*<br>#*Features*: *f* | $p \times f$<br>#*LSTMs*: *p*<br>#*Features*: *f* |
| Layer 1 | LSTM, 128 units,<br>activation = relu<br>Sequence- to- sequence | LSTM, 128 units,<br>activation = relu<br>Sequence- to- sequence | LSTM, 128, units,<br>activation = relu<br>Sequence- to- sequence |
| Layer 2 | LSTM, 128, units,<br>activation = relu<br>Sequence- to- one | LSTM, 128, units,<br>activation = relu<br>Sequence- to- one | LSTM, 128, units,<br>activation = relu<br>Sequence- to- one |
| Layer 3 | Dense, 3 units,<br>activation = softmax<br>outputs = 3 classes | Dense, 32 units,<br>activation = relu | Dense, 32 units,<br>activation = relu |
| Layer 4 | × | Dense, 3 units,<br>activation = Softmax<br>Output = 3 classes | Dense, 3 units,<br>activation = Softmax<br>Output = 3 classes |

https://doi.org/10.1371/journal.pone.0274516.t001

occurrences are calculated in 13 directions, resulting in a feature map of size $13 \times h \times w$ for every VS. The feature maps from all VSs are stacked and fed to LSTM-based deep neural architecture for fusion after vectorization and padding. The architecture consists of three LSTM and two dense layers. The output dense layer contains three units with a softmax activation function for classification. The Relu activation function is used in the other layers. Table 1 displays the number of essential parameters used in different layers of the proposed method for all of the input modes.

### Dataset

The LIDC-IDRI dataset proposed by Armato et al. [60] is applied to assess these methods. This dataset embodies the analytical CT scans of the lung, designing a CAD system to detect and classify lung carcinoma. There are 1018 cases in this dataset, collected and compiled through contributions from several medical imaging institutes and academic hubs. Each subject's case includes CT scan images and their associated XML files representing the annotations from four radiologists. The following three nodule sizes show the classification criteria for the lesions.

- Nodule: diameter $\geq$ 3mm

- Nodule: diameter $\leq$ 3mm

- Non-nodule: diameter $\geq$ 3mm

The primary task of a radiologist is to discover every possible lung nodule in the entire CT scan images. The CT scan images applied in this study are arranged in a sequence of slices for every patient; each image is labeled as malignant, benign, or ambiguous.

Through preprocessing, we found 796 nodules with a diameter of $\geq$ 3mm in the selected CT scans. The initial 300 cases are selected for the experiments. In this study, some cases do not have nodules, and some nodules are smaller than 3mm in diameter, indicating a low probability of malignancy. After removing the CT scans with inconsistent slice spaces and clean CT images (i.e., CT scans with no nodules), 779 nodules were selected for experiments. The method proposed by Naik et al. [5] was adopted to categorize nodules classes. In this study, one slice from each nodule (779) is selected randomly to assess the 2D-GLCM model and WSA-Otsu segmentation. The training and test data constitute 80% and 20% of the data, respectively. All the slices are first converted into grey-scale intensities in the interval (0, 255).

LUNGx is another dataset used in the experiments and is associated with the 2015 SPIE medical imaging symposium challenge [61]. It contains 60 test scans containing 73 nodules, 37 of which are benign and 36 are malignant. According to the challenge guidelines, we followed the suggestion of using the LIDC-IDRI dataset for training the models, and only evaluated the performance of our model on the testing set provided by LUNGx dataset.

## Results and discussion

### Evaluation criteria

Lung cancer diagnosis is a classification procedure where the CT scan images are classified as malignant, benign, and ambiguous. Most available studies have focused on lung cancer classification using indicators such as accuracy, sensitivity, and specificity to determine the efficiency of the methods. In addition to these indicators, in this study, a multi-class one-vs-rest (OvR) algorithm is used to calculate the ROC AUC averages for the three classes. Moreover, the macro and micro averaging methods are used for experimental comparison, and the ten-fold cross-validation is applied on the training and test data.

### Detection results

The experimental performance of the proposed WSA-Otsu segmentation is compared with other comparative thresholding methods. In this comparison, 779 CT scans covering nodules with different types, shapes, and sizes are selected. The detection accuracy is calculated based on the reference masks in the dataset. The comparison between WSA-Otsu and the latest proposed thresholding methods is shown in Table 2. As can be seen, WSA-Otsu outperforms other methods in nodules detection. The highest F-score value is obtained by the proposed methods.

The significance of hybridization between WSA and the Otsu thresholding method can be seen in Fig 5, which shows the fitness function convergence curves for WSA and other metaheuristic algorithms based on Otsu thresholding for ten images and five levels. WSA-Otsu method achieves a higher convergence rate than other Otsu-based metaheuristic methods, which can be attributed to the global search capability and efficiency of WSA.

Fig 6 shows the thresholding time of the WSA-Otsu and the latest proposed methods, where two, three, four, and five thresholds are applied for detecting nodules in the lung CT scan images. The algorithms execution time increases when more thresholds are considered. This is due to increased dimensionality of the thresholds optimization problem, effectively reducing the speed of segmentation process. Exploiting WSA to find the proper thresholds in the proposed method reduces the thresholding execution time. The diagram analysis in Fig 6

**Table 2. The comparison of WSA-Otsu with counterpart methods.**

| Methods | TP | TN | FP | FN | F1-Score |
|---|---|---|---|---|---|
| Standard Otsu method [16] | 95 | 38 | 5 | 17 | 89.6% |
| Otsu based Darwinian particle swarm optimization (DPSO) [62] | 89 | 49 | 7 | 10 | 91.3% |
| Statistical and Shape-Based Features Lung nodules detection [63] | 101 | 32 | 6 | 16 | 90.2% |
| Otsu based exchange market algorithm (EMA) [64] | 93 | 48 | 11 | 3 | 93.0% |
| Adaptive Particle Swarm Optimization– Gaussian mixture model (APSO-GMM) [65] | 109 | 33 | 8 | 5 | 94.4% |
| FFNN + combination of cuckoo and PSO [66] | 112 | 33 | 6 | 4 | 95.7% |
| **WSA-Otsu** | **106** | **42** | **3** | **4** | **96.8%** |

https://doi.org/10.1371/journal.pone.0274516.t002

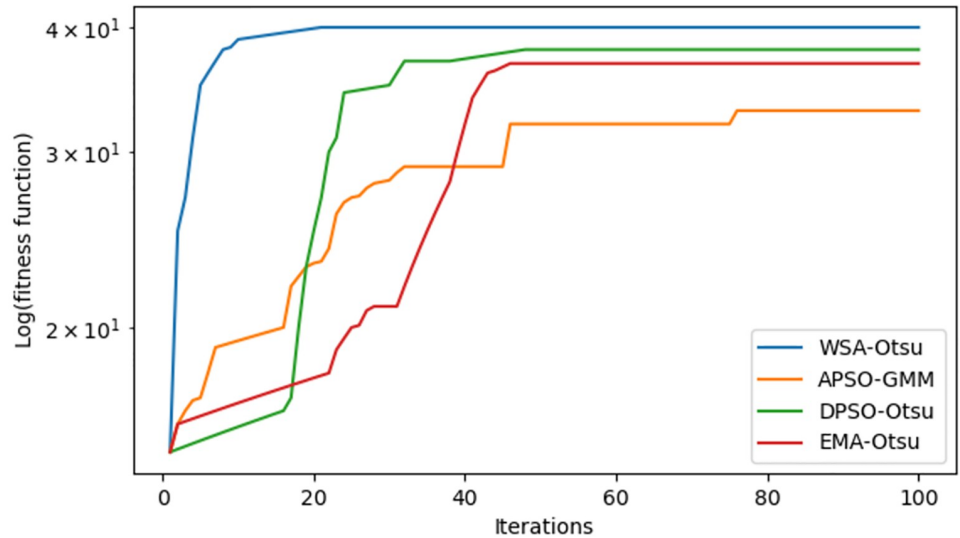

**Fig 5. The convergence curves of Otsu method in terms of fitness values at five different levels.**

https://doi.org/10.1371/journal.pone.0274516.g005

indicates that as the number of thresholds increases, the thresholding time in the competitive thresholding methods becomes considerably longer than that of the proposed method. The assessment of the acceleration and speed of the segmentation process reveals that when the number of thresholds is five, the proposed method finds thresholds 9.1 times faster than the Otsu method and 2.3 times faster than the DPSO-Otsu method.

## Classification results

Three GLCM computation modes, explained earlier, are used to extract the texture features provided to the proposed LSTM-based deep fusion structure for classification. The fusion

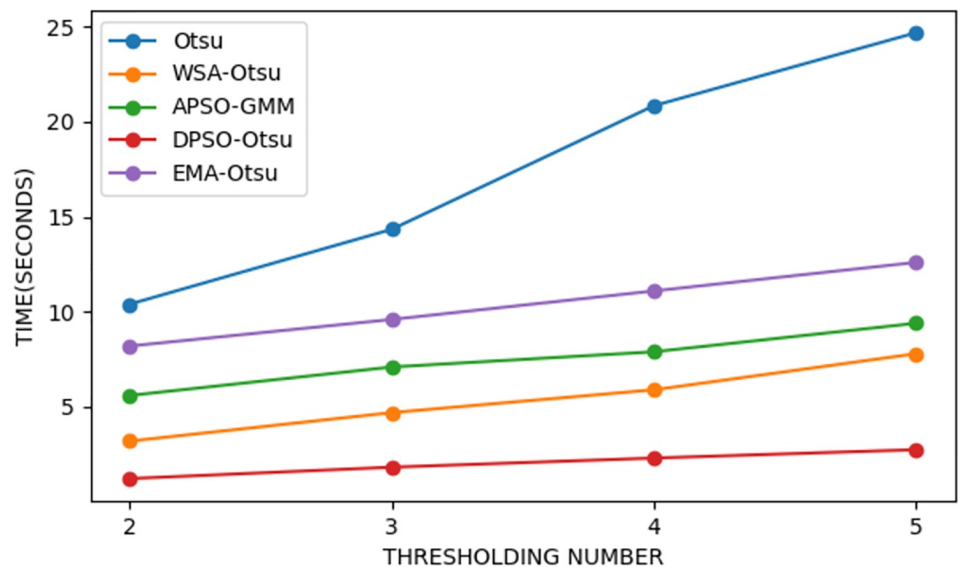

**Fig 6. The thresholding time of the proposed method vs. other counterparts.**

https://doi.org/10.1371/journal.pone.0274516.g006

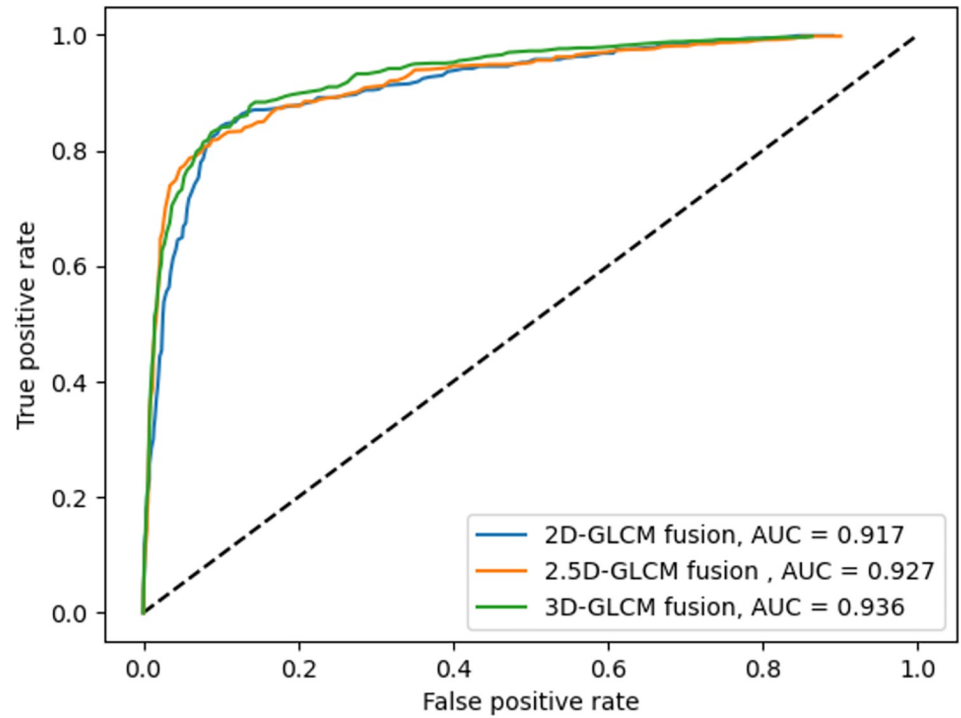

**Fig 7. ROC curves for three fusion models and their corresponding macro-averaged AUC values.**

https://doi.org/10.1371/journal.pone.0274516.g007

network for 2D-GLCMs is evaluated with the data from 779 slices selected from the dataset of different patients, from which 155 are reserved for the testing phase. Every slice is labeled as either benign, malignant, or ambiguous, depending on their respective nodule. The ROC curves are plotted in Fig 7, including the macro-average AUC, where the metric is calculated for each class independently, and their average is then obtained. As there are many ambiguous subject samples within the dataset, allocating an equal weight to all classes in averaging may prevail the ambiguous class over other classes. Therefore, the micro-averaged AUC, where the impact of each class is aggregated for averaging is also shown in Fig 8. When the fusion architecture is applied on 2D-GLCMs, the final test results on the accuracy, sensitivity, and specificity are respectively 94.4%, 91.6%, and 95.8%.

The 300 CT scans containing 779 nodules with diameter ≥ 3mm are selected for the evaluation of the fusion architecture on 2.5D-GLCMs, from which 155 nodules are used as test set. The accuracy, sensitivity, and specificity obtained in this mode are respectively 97.33%, 96%, and 98%. Similarly, for the same 155 test nodules, the accuracy, sensitivity, and specificity obtained for nodule classification using the proposed LSTM-based deep fusion model for 3D-GLCMs computed from VSs are 98.7%, 98%, and 99%, respectively.

This study also aims to investigate the effectiveness of the proposed deep fusion of texture features, computed using different modes of GLCMs, in comparison with other CNN-based deep lung nodule classification models. Different CNN schemes are implemented and evaluated on the LIDC-IDRI dataset to confirm the potential of the proposed method in the experiments. MobileNetV2 [67], EfficientNet-B5 [68], and 3D-CNN [50] are trained through the raw CT images. In this study, the last layers of the networks are the output of the three-class nodules (Table 3). As observed, the LSTM-based deep fusion structure for 2D-GLCMs outperforms MobileNetV2 and EfficientNet-B5. However, its performance is lower than the

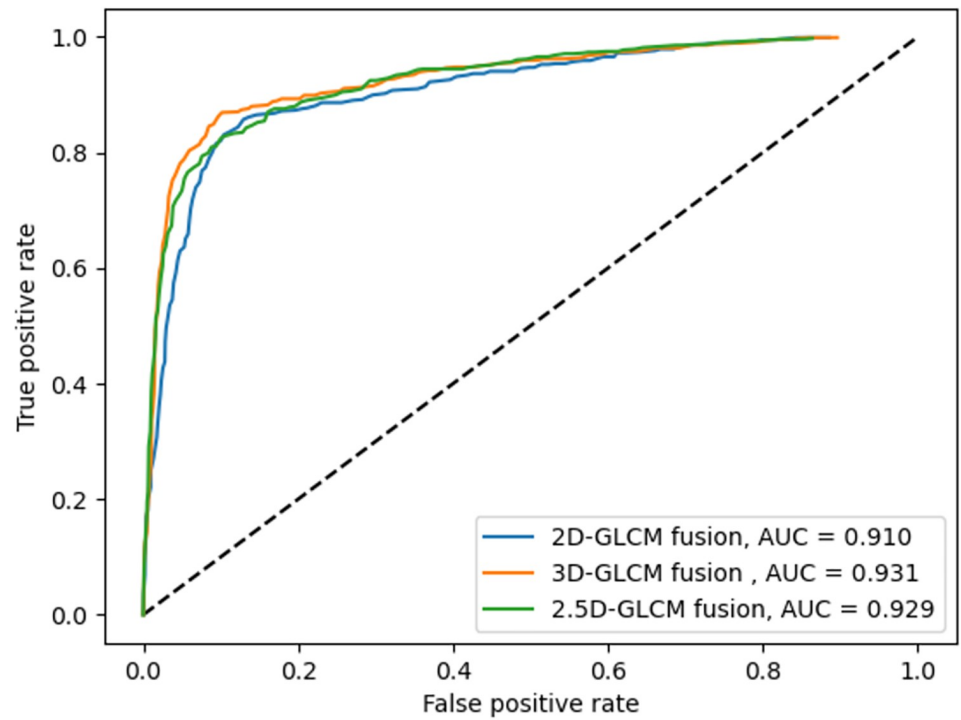

**Fig 8. ROC curves for three fusion models and their corresponding micro-averaged AUC values.**

https://doi.org/10.1371/journal.pone.0274516.g008

3D-CNN trained with multiple raw 2D slices. When using 2.5D-GLCMs, extracted from multiple 2D-slices of a nodule and presented to the proposed LSTM-based deep fusion structure, the model outperforms 3D-CNN in nodule classification. These results acknowledge the importance of volumetric appearance of lung nodules for improving the model's effectiveness in classification.

Table 4 compares the performance of the proposed method with the latest alternative approaches for nodule classification, tested on the LIDC-IDRI dataset. Early CAD systems employed traditional machine learning methods for classification and thus their performance are usually inferior to deep methods. For example, Iii et al. [71] applies linear discriminant analysis, achieving 70% sensitivity, Suzuki et al. [62] applied an artificial neural network, resulting in 80.3% sensitivity, and Teramoto and Fujita [63] used cylindrical filters with an SVM to detect and classify NSCLC, and showed 80% sensitivity.

**Table 3. A comparison between the proposed 2D and 2.5D GLCM fusion methods and other deep CNN-based methods regarding the accuracy, sensitivity, and specificity metrics.**

| Method | Accuracy ±*std*(%) | Sensitivity ±*std*(%) | Specificity ±*std*(%) |
|---|---|---|---|
| MobileNetV2 [67] | 82.97 ±1.3 | 69.37 ±.8 | 92.12 ±1.1 |
| EfficientNet-B5 [69] | 88.77±1.7 | 94.59±0.2 | 84.85±0.5 |
| ResNet50 [70] | 86.23±2.1 | 98.2±0.2 | 78.18±1.02 |
| Xception [71] | 92.39±1.2 | 93.69±1.55 | 91.52±0.9 |
| NASNetMobile [62] | 87.68±0.22 | 74.77±0.6 | 96.36±0.56 |
| Raw CTs-based 3D-CNN [50] | 97.17±0.25 | 87±0.57 | 94±0.11 |
| LSTM-based deep fusion of 2D-GLCMs | 94.4±0.7 | 91.6±0.32 | 95.6±0.58 |
| **LSTM-based deep fusion of 2.5D-GLCMs** | **97.33±0.9** | **96±1.6** | **98±0.77** |

https://doi.org/10.1371/journal.pone.0274516.t003

Table 4. A comparison between the proposed fusion methods for different modes of GLCM computation and the recently proposed classification methods regarding the accuracy, sensitivity, and specificity metrics.

| Research | Method | Accuracy | Sensitivity | Specificity |
|---|---|---|---|---|
| [65] | Linear Discriminant analysis | × | 70% | 78% |
| [62] | MTANN | × | 80.30% | 83.5% |
| [72] | FLD classifier | × | 82.66 | 84.6% |
| [73] | a neural classifier | × | 87.50% | 89.3% |
| [63] | cylindrical filters and SVM | × | 80% | 88.6% |
| [74] | Hierarchical Vector Quantization and SVM | × | 82.70% | 86.3% |
| [75] | GLMR classifier | × | 92.91% | 89.3% |
| [76] | LDA classifier and optimal thresholding | 84% | 97.14% | 80.2% |
| [77] | backpropagation network | 90.70% | × | 87.7% |
| [78] | fuzzy inference method | 94.12% | × | 96% |
| [79] | texture and learned distance metrics and TSCBIR classifier | 91% | × | 81.1% |
| [80] | probabilistic neural network | 92% | 95% | 90% |
| [81] | multilayer feed-forward neural network with supervised learning method as a classifier | 95% | 100% | 93.5% |
| [34] | double-path convolutional neural network (DPCNN) | 70.4% | × | 77.8% |
| [37] | local energy-based shape histograms + Adaboost | 84.6% | 71.4% | 100% |
| [44] | 3D-CNN-based scans. | 83.33% | × | 94% |
| [51] | GLCM based CNN-RNN | 76% | × | 84.5% |
| [48] | 3D-GLCM-CNN | 93% | 90% | 89% |
| [82] | 3D-convolution-LSTM | 97.2% | 98.2% | 92.6% |
| [45] | CNN-LSTM | 97% | × | 91.7% |
| [83] | Fusion of clinical and CT-scan-based features | 93.6% | 91.9% | 95.6% |
| This study | **LSTM-based deep fusion of 2D-GLCMs (Proposed)** | **94.4±0.7%** | **91.6±0.32%** | **95.6±0.58%** |
| This study | **LSTM-based deep fusion of 2.5D-GLCMs (Proposed)** | **97.33±0.9%** | **96±1.6%** | **98±0.66%** |
| This study | **LSTM-based deep fusion of 3D-GLCMs (Proposed)** | **98.7±0.55%** | **98±0.4%** | **99±0.98%** |

https://doi.org/10.1371/journal.pone.0274516.t004

The results presented in Table 4 indicate the effectiveness of the LSTM-based deep fusion structure applied on the stack of 3D-GLCMs extracted from VSs of nodules. Instead of performing 3D-GLCM calculations using on the whole CT scan volume like Deepti [39], in this study the 3D-GLCM calculations are confined to the VSs. This not only decrease the calculation costs, but fusing these sequential features in the proposed LSTM-based deep structure leads to a superior performance to similar methods like that of Tan et al. [48]. It also reveals that the higher-level features learnt in the deep structure from the valuable information extracted through GLCM computations provide higher accuracy than direct processing of raw CT scan image data in other end-to-end methods, like that of Puttagunta and Ravi [64]. It should be noted that such advantage over utilizing raw CT images in lung cancer classification is observed even with the fusion of 2.5D-GLCMs.

## Baseline models

The two main components of the proposed model, namely stacked 3D-GLCMs obtained from VSs and the LSTM-based deep fusion architecture are investigated in more detail. This assessment emphasizes the functional gains of the proposed method over other baseline models like [48, 84] in nodule classification. Figs 9 and 10 show the ROC curves for these models and their corresponding micro- and macro-averaged AUCs, respectively. The best micro- and macro-average AUCs were obtained by the proposed LSTM-based fusion of 3D-GLCM information. The second best AUC was obtained when the in-depth features were extracted through

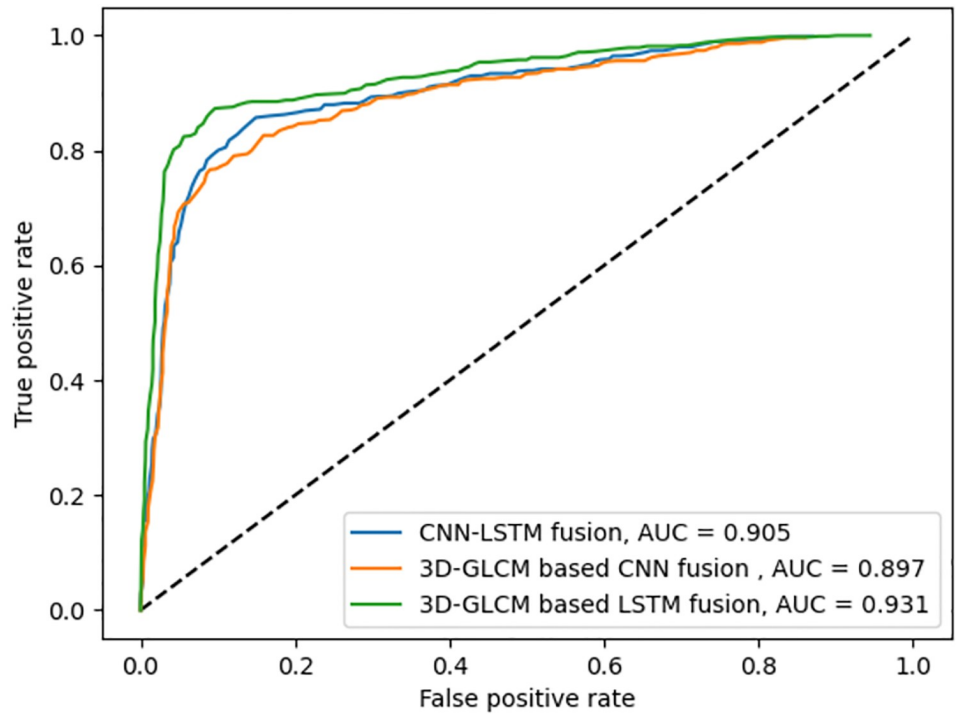

**Fig 9. ROC curves for baseline and proposed models, and their corresponding micro-averaged AUC values.**

https://doi.org/10.1371/journal.pone.0274516.g009

convolutional layers and fused by an LSTM network. The results indicate that the computation of 3D-GLCMs within VSs allows more spatial relationships between nodule regions to be captured, and the LSTM-based deep fusion can model the long-range dependencies between these regions of the nodule volume. Lower fusion impact is observed when fusing 3D-GLCMs through a CNN model, demonstrating the importance of considering long-range dependencies in lung nodules by recurrent neural networks.

The performance of the proposed method is also compared with the baseline models on the LUNGx dataset. The results are tabulated in Table 5. As no training data is available in the LUNGx dataset, the acquired models are only tested on the supplied testing data. Such an experiment provides a thorough evaluation of the models robustness and generalizability.

Moreover, we run a T-test to investigate the statistical significance in the difference of the performance results obtained by the proposed model compared with other baseline models. Our alternative hypothesis is the higher average AUC of the LSTM-based deep fusion model applied on 3D-GLCMs in comparison to the baseline models. The *p-values* obtained are tabulated in Table 6. These *p-values* confirm our alternative hypothesis that the proposed fusion architecture for stack of 3D-GLCMs is statistically better than the other possible alternatives.

## Ablation study

Two experiments are conducted in this section to exhibit the importance of the features fed to the proposed model. In the first experiment, instead of directly feeding the 3D-GLCMs to the fusion model, 16 texture feature descriptors [28] are calculated from these 3D-GLCMs and then fed to the proposed LSTM-based deep fusion structure. In the second, the 3D-GLCMs computed from different VSs in each direction are stacked together and fed to the LSTM-

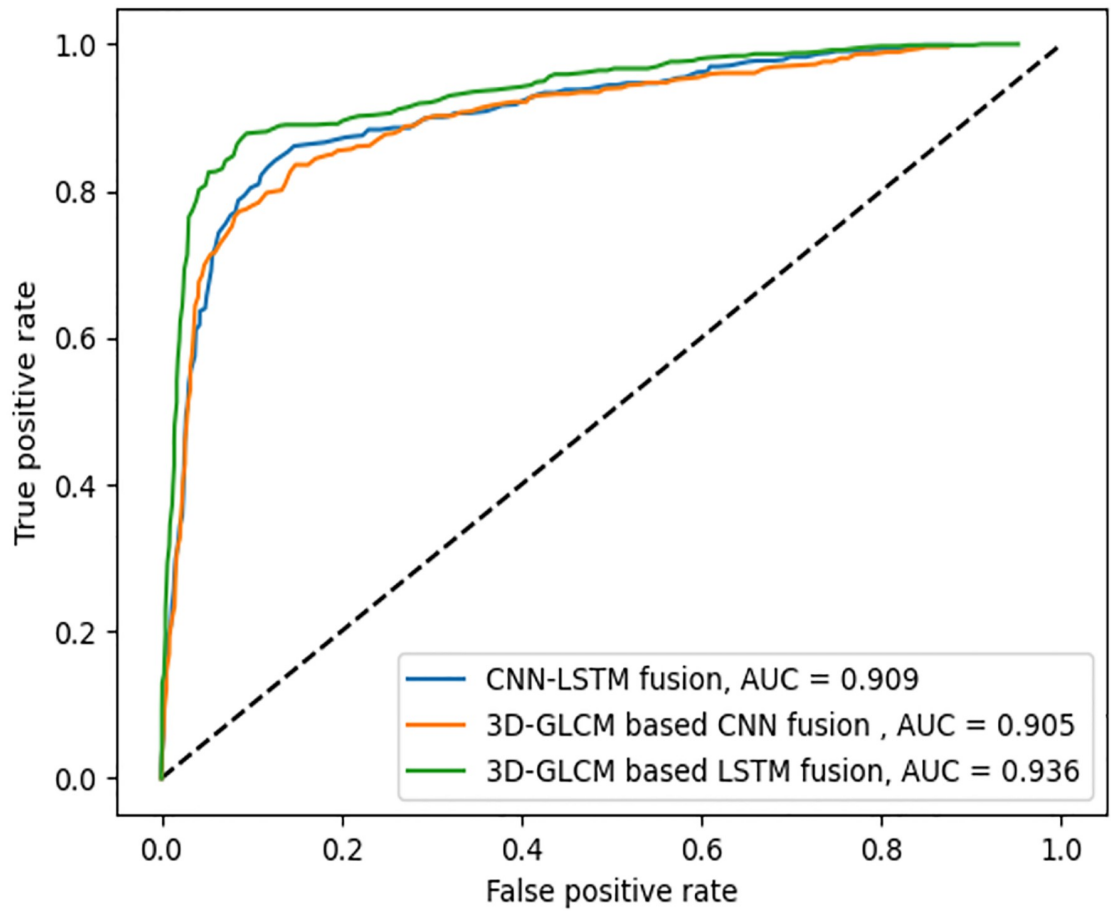

**Fig 10. ROC curves for baseline and proposed models, and their corresponding macro-averaged AUC values.**

https://doi.org/10.1371/journal.pone.0274516.g010

**Table 5. The classification results in terms of micro-averaged AUCs trained on the LIDC-IDRI dataset and tested on LIDC-IDRI and LUNGx datasets for the proposed and baseline models.**

| models | AUCs±*std* (%) | |
| --- | --- | --- |
| | **LUNGx dataset** | **LIDC-IDRI dataset** |
| CNN-LSTM fusion | 51.3±7.1 | 90.5±1.9 |
| 3D GLCM based CNN fusion | 56.05±5.1 | 89.7 ±2.4 |
| 2.5D GLCM based LSTM fusion | 61.3±4.9 | 92.9±2.6 |
| 3D GLCM based LSTM fusion | 70.02±3.5 | 93.1±1.8 |

https://doi.org/10.1371/journal.pone.0274516.t005

**Table 6. The T-test *p*-values obtained comparing the average AUC of the LSTM-based deep fusion of 3D-GLCMs with other baseline models.**

| models | *p-value* |
| --- | --- |
| CNN-LSTM fusion | 0.0017 |
| 3D GLCM based CNN fusion | 0.000018 |
| 2.5D GLCM based LSTM fusion | 0.02 |

https://doi.org/10.1371/journal.pone.0274516.t006

**Table 7. Results of the ablation study for the type of features presented to the LSTM-based deep fusion model.**

| Features used | Accuracy | Sensitivity | Specificity |
| --- | --- | --- | --- |
| 3D-GLCMs stacked for each VS | 98.7% | 98% | 99% |
| Texture feature descriptors | 91.3% | 89.3% | 90.3% |
| 3D-GLCMs stacked for each direction | 92.7% | 91.2% | 88.3% |

https://doi.org/10.1371/journal.pone.0274516.t007

based deep fusion structure with 13 LSTM units, corresponding to each of the 13 possible directions, to combine the texture features extracted for each direction (Table 7).

The results show that the fusion of feature descriptors instead of 3D-GLCMs, decrease accuracy, sensitivity, and specificity of nodule classification. Moreover, stacking the 3D-GLCMs for each direction before feeding to the proposed fusion architecture negatively affects the model's classification performance. The first experiment results can be assigned to valuable information loss during the estimation of the texture feature descriptors, and the second experiment results can be attributed to the uncorrelated GLCM directions.

## Computational complexity

This section presents the time complexity estimation of the proposed model. LSTM is a local model in space and time [33] and the input size does not affect the storage requirement. Thus the time and space complexity of each LSTM unit (e.g. for one time-step) is $O(d)$ and $O(1)$, respectively, where $d$ is the input size. Let $n$ slices in each nodule volume be grouped into $p$ VSs ($p < n$) so that each element of the VS has a resolution of $1mm^3$. Then, fusing 2.5D-GLCMs and 3D-GLCMs in the proposed architectures would have a complexity of $O(nd)$ and $O(pd)$, respectively. When repeating the training process of these models for $i$ epochs the complexities are respectively increased to $O(ind)$ and $O(ipd)$, ignoring the overhead for the final classification.

The long-range dependencies within volumetric nodules can be captured through the CNN-LSTM-based fusion models, where convolutional layers are applied to each slice to extract deep features, fed next to LSTM units. The 3D-GLCMs were also fused through CNN model [48]. However, these approaches have been shown to increase the overall computational complexity [84]. A comparison of the time requirement in both train and test phases of different models is presented in Figs 11 and 12. The training results, separately depicted for five epochs, confirm that the proposed 3D-GLCM model consumes less cumulative training time than the other two models. Similar results can be seen for testing time. These results meet the hypothesis of reducing the complexity when considering the long-range dependencies in 3D nodule volumes by breaking them down to VSs.

## Conclusions and future work

Discrimination of malignant and benign tumors is one of the most challenging tasks of physicians in analyzing CT scan images. The present study proposed a 3D-GLCM computation approach based on VSs, fed to an LSTM-based deep fusion structure to learn more valuable representation features, a task shown to be difficult for CNN-based models. In the proposed method, 3D-GLCMs are computed from VSs of a CT scan to represent more valuable tissue texture data. An LSTM-based deep fusion of adjacent VSs features is used for classifying lung nodules based on the spatially correlated information between the sequence of VSs.

Most researchers used machine learning models to classify and detect lung cancer based on CT scan images. Nowadays, deep learning methods like CNNs prevail in the same context.

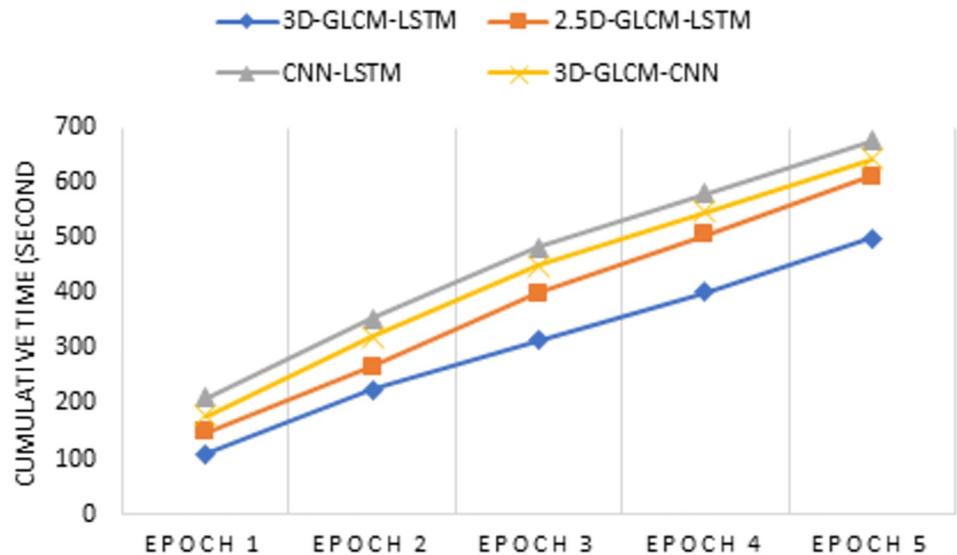

**Fig 11. The training time of different models for five epochs.**

https://doi.org/10.1371/journal.pone.0274516.g011

Employing 3D-GLCM features fused in an LSTM-based deep fusion structure is more accurate than providing 3D-GLCM to a CNN model presented in other studies [37].

Experimentally, LSTM has demonstrated better performance in modeling the irregularity of sequential data patterns in CT scan slices. GLCM is calculated in different modes, where each slice is first considered separately and classified into begin, malignant, and ambiguous, and then, the LSTM-based deep fusion structure fuses these features from different directions (2D-GLCM). LSTM boasts higher accuracy than other deep transfer CNN methods such as MobileNetV2, EfficientNet-B5, and ResNet50, which consider a single slice as the input. The features are extracted from the sequenced slices and fed to the LSTM (2.5D-GLCM), achieving higher accuracy than the 2D-GLCM mode or providing GLCM to a CNN fusion model [37].

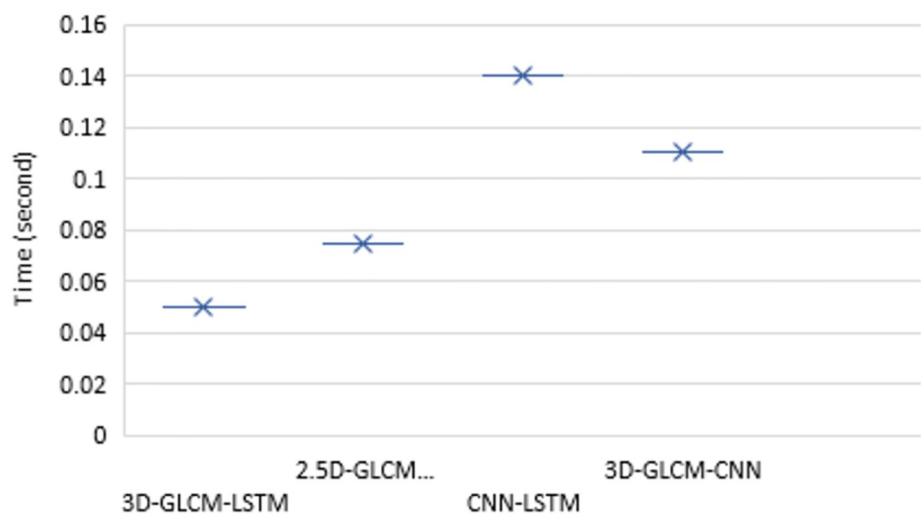

**Fig 12. The test time of different models.**

https://doi.org/10.1371/journal.pone.0274516.g012

The significance of calculating 3D-GLCM from adjacent VSs is experimentally demonstrated. Fusing these features into an LSTM-based deep fusion structure achieves the highest accuracy.

The advantage of the improved Otsu method is that it requires less execution time than the conventional Otsu, especially at a high threshold count, and increases the neural network learning speed. Based on the experimental results, the proposed WSA-Otsu method contributes highly to lung nodules detection. The thresholding time is 9.1 times lower than the standard Otsu and 2.3 times lower than the Otsu-based DPSO heuristic algorithm. The results of applying the proposed WSA-Otsu and other thresholding methods on medical images are shown in Table 2, where, as observed, WSA-Otsu has the highest score.

Lung nodule texture features can be extracted with different methods. It is possible to increase the efficiency of features significantly. Fusing features such as the local binary patterns, shape index histograms, local energy patterns, and grey level length matrices can accomplish this objective.

Another research topic can be combining similar elements in a feature fusion framework to learn the adopted features close to each other. Using the newly developed approach known as the attention mechanism focusing on high-level features, the researchers plan to conduct a study on avoiding irrelevant or redundant features in calculating GLCMs from different directions via the attention mechanism. The non-local operations can be designed to extract the global and local high-level features of the lung nodules.

## Author Contributions

**Conceptualization:** Ahmed Saihood, Hossein Karshenas.

**Data curation:** Ahmed Saihood.

**Formal analysis:** Ahmed Saihood, Hossein Karshenas.

**Methodology:** Ahmed Saihood, Hossein Karshenas.

**Project administration:** Hossein Karshenas.

**Software:** Ahmed Saihood.

**Supervision:** Hossein Karshenas.

**Validation:** Ahmed Saihood, Ahmad Reza Naghsh Nilchi.

**Visualization:** Ahmed Saihood.

**Writing – original draft:** Ahmed Saihood.

**Writing – review & editing:** Hossein Karshenas, Ahmad Reza Naghsh Nilchi.